# Materials Science & Engineering A

# A novel and sustainable method to develop non-equiatomic CoCrFeNiMox high entropy alloys via spark plasma sintering using commercial commodity powders and evaluation of its mechanical behaviour

--Manuscript Draft--

| | |
|---|---|
| **Manuscript Number:** | MSEA-D-23-00557 |
| **Article Type:** | Research Paper |
| **Keywords:** | High entropy alloys, spark plasma sintering, sustainability, commercial commodity powders |
| **Corresponding Author:** | Venkatesh Kumaran S<br>IMDEA Materials Institute<br>Getafe, Madrid SPAIN |
| **First Author:** | S Venkatesh Kumaran |
| **Order of Authors:** | S Venkatesh Kumaran |
| | Dariusz Garbiec |
| | José Manuel Torralba |
| **Abstract:** | A novel approach to developing high entropy alloys (HEAs) using spark plasma sintering (SPS) was explored in this work where a mix of commercial commodity powders like Ni625, CoCrF75, and 316L was used instead of pre-alloyed powders avoiding the expensive pre-alloying steps like mechanical alloying or gas atomizing. Three non-equiatomic HEAs, based on Co, Cr, Fe, Ni, and Mo were designed and developed by blending the powders which were sintered via SPS and resulted in a single FCC phase after homogenization. The HEAs were microstructurally and mechanically characterized with tensile and hot compression tests up to a temperature of 750oC showing excellent properties. The maximum room temperature tensile strength and ductility demonstrated was 712 MPa and 62% respectively, by the alloy Co23.28Cr28.57Fe25.03Ni21.01Mo2.1. Moreover, the same alloy exhibited a compression strength greater than 640 MPa with a ductility above 45% at a temperature of 750oC. Also, this study paves the way for a novel fabrication route that offers more flexibility to develop new HEAs cost-effectively and efficiently which is crucial for the discovery of new materials over high-throughput techniques. Using such commodity alloys also opens the possibility of developing ingot casting from recycled scraps avoiding the direct use of critical metals. |
| **Suggested Reviewers:** | Laura Cordova Gonzalez<br>laura.cordova@chalmers.se<br>Expert in powder rheology, sustainable manufacturing, sintering |
| | Sheng Guo<br>sheng.guo@chalmers.se<br>Expert in high entropy alloys |
| | Eduard Hryha<br>hryha@chalmers.se<br>Expert in powder processing, sintering |

# A novel and sustainable method to develop non-equiatomic CoCrFeNiMo$_x$ high entropy alloys via spark plasma sintering using commercial commodity powders and evaluation of its mechanical behaviour

S Venkatesh Kumaran[1,2], Dariusz Garbiec[3], José Manuel Torralba[1,2]

1. IMDEA Materials Institute, Madrid, Spain, 28906
2. Universidad Carlos III de Madrid, Leganes, Spain, 28911
3. Metal Forming Institute, 14 Jana Pawla II Street, 61-139 Poznan, Poland

## 1. Abstract

A novel approach to developing high entropy alloys (HEAs) using spark plasma sintering (SPS) was explored in this work where a mix of commercial commodity powders like Ni625, CoCrF75, and 316L was used instead of pre-alloyed powders avoiding the expensive pre-alloying steps like mechanical alloying or gas atomizing. Three non-equiatomic HEAs, based on Co, Cr, Fe, Ni, and Mo were designed and developed by blending the powders which were sintered via SPS and resulted in a single FCC phase after homogenization. The HEAs were microstructurally and mechanically characterized with tensile and hot compression tests up to a temperature of 750°C showing excellent properties. The maximum room temperature tensile strength and ductility demonstrated was 712 MPa and 62% respectively, by the alloy $Co_{23.28}Cr_{28.57}Fe_{25.03}Ni_{21.01}Mo_{2.1}$. Moreover, the same alloy exhibited a compression strength greater than 640 MPa with a ductility above 45% at a temperature of 750°C. Also, this study paves the way for a novel fabrication route that offers more flexibility to develop new HEAs cost-effectively and efficiently which is crucial for the discovery of new materials over high-throughput techniques. Using such commodity alloys also opens the possibility of developing ingot casting from recycled scraps avoiding the direct use of critical metals.

## 2. Introduction

High Entropy Alloys (HEAs) have garnered significant research interest in the past decade since the advent of cantor and cantor based alloys [1], [2], due to their unconventional alloying approach which results in exceptional mechanical properties, such as high strength [3], [4], wear resistance [5], corrosion resistance [6]–[8], and thermal stability [9] among others. Despite ingot metallurgy being the most common method to manufacture HEAs, powder metallurgy (PM) has also shown great potential in manufacturing HEAs. This is mainly because PM can achieve significantly higher compositional accuracy, prevent segregation, and achieve superior microstructural control [10]. Powder metallurgy high-entropy alloys (PMHEAs) have been developed using three different classes of powders to date: fully pre-alloyed gas-atomized powders, pure elemental powders, and fully pre-alloyed mechanically alloyed powders (which are also fabricated from elemental powders). However, some of the main limitations in PM routes to fabricate HEAs are the cost and availability of the powders. Firstly, there are no fully pre-alloyed powders in the market to fabricate even the most extensively used HEAs today. Thus, there is always a need to start from expensive elemental powders (>99% purity) to proceed with either mechanical alloying to produce pre-alloyed powders or by using critical metals as raw materials to produce powders by gas atomizing, both of which are time-consuming and expensive. Moreover, pure elemental powders can also be difficult to handle. For instance, in the commonly studied cantor alloy (CoCrFeMnNi), Cr has a high affinity towards oxygen forming chromium oxides, whereas, Ni and Co are considered hazardous by the REACH regulations [11], [12]. However, as described in our previous work [13], there are many grades of powder available in the market that belong to the families of metals on which many HEAs are based: Ni, Cr, Fe, Co, Ti, Al, etc. These grades are all commercially available from several manufacturers and can be delivered in large quantities within short times at competitive prices. So, in this work, we explore the possibility of using these families of available alloys in mass production, designated by the term 'commodity', as raw materials to develop HEAs. As these commodity powders are usually mass-produced, their cost is relatively low and are easily available in the market, due to which, they offer a highly convenient way for the development of engineering materials by the PM route.

In our previous work [13], the focus was to develop a prototype model using field-assisted hot pressing to analyze the microstructure which proved that developing HEAs using commercial commodity powders was very much feasible. The composition was designed based on calculating the thermodynamic parameters to ensure a single solid solution as detailed in our previous work [13]. In our current work, we developed three different non-equiatomic HEAs using the available commercial commodity powders using Spark Plasma Sintering (SPS). Tensile and hot compression tests were performed to evaluate the mechanical behavior. The alloys developed were $Co_{25.56}Cr_{20.7}Fe_{26.92}Ni_{25.2}Mo_{1.6}$, $Co_{23.28}Cr_{28.57}Fe_{25.03}Ni_{21.01}Mo_{2.1}$, $Co_{29.07}Cr_{17.2}Fe_{28.22}Ni_{24.45}Mo_{1.06}$, which were labeled as C1, C2, and C3 respectively. Today, the use of critical metals is controversial, and this can be an obstacle in developing HEAs as many of them contain Co and Ni. So, with this work, a new possibility is opened to manufacture HEAs both by casting and powder metallurgy with commodity alloy powders or scraps as raw material and thus avoiding the direct use of critical metals.

Recently, single-phase FCC-type HEAs have been widely studied, such as CoCrFeNi alloy due to their exceptional ductility and fracture toughness [14]. However, they possess lower strengths at ambient and elevated temperatures making them unsuitable for structural applications. In this regard, many alloying elements have been added to the CoCrFeNi system to improve the mechanical properties. Specifically, Mo improves the room and high-temperature strength owing to its high melting point and elastic modulus [15]. So, motivated by this, Mo was included through the commodity powders to develop three different alloys with varying Mo content and the mechanical properties were tested both at room and at high temperatures.

## 3. Materials and methods

In this work, three non-equiatomic $CoCrFeNiMo_x$ alloys were prepared using commercial commodity powders. The commodity powders used were Ni 625, SS 316L, Invar 36, CoCr F75, and Fe49Ni. The CoCr alloy powders were provided by VDM Metals (Germany) and the rest were from Sandvick Osprey (UK). All the powders used were commercial grades, ready to be used for other applications. The particle size distribution and the composition of the powders are listed in Table 1. From these above powders, three different alloy combinations were designed and were simply mixed in the appropriate proportion as shown in Table 2. The alloys are labeled C1, C2, and C3. The composition was designed to keep Ni, Fe, Cr, and Co fairly equiatomic with minor Mo addition. The morphology of the C1, C2, and C3 powder mixes are shown in Fig. 1. In C1, as shown in Fig. *1* (a) and (b), Ni 625 consists of an irregular shape with a d50 of 60µm, which is much larger than the other powder particles. These powders are relatively spherical, but due to a non-optimized gas atomization process, it has led to clusters of satellites. As seen in Fig. *1* (d), the 316L powder particles in the C2 mix are adhered to each other forming large agglomerates. This is attributed to the smaller particle size distribution (PSD) of 316L. In C3, the Fe49Ni particles are highly irregular and adhered to each other and also to the CoCrF75 powder particles as shown in Fig. *1* (e). Once the powders were mixed, the SPS consolidation was carried out in HP D 25/3 SPS furnace (FCT system, Germany) at Siec Badawcza Lukasiewicz, Poznan, Poland. A tungsten foil was placed between the powders and the graphite die to avoid carbon diffusion. The die had a cylindrical shape with 60 mm diameter and 15 mm height. The sintering was carried out at two different temperatures for each alloy, 1000°C, and 1100°C with a heating rate of 100°C/min with a 50 MPa pressure for 10 minutes. Initially, based on the previous works in the literature [10], the temperatures were chosen to be 1000°C and 1250°C. When C1 was being consolidated, a liquid phase was formed at around 1140°C and removed from the tools due to which the consolidation temperature for C2 and C3 was reduced to 1100°C. After sintering, the bulk samples were annealed at 1200°C for 24 hours in a metal muffle furnace (Carbolite CWF 13/23) to homogenize the microstructure. All the samples were vacuum encapsulated in a glass tube to avoid oxidation. Both the as-sintered (AS) and heat-treated (HT) samples were manually ground and polished up to a colloidal silica solution of 0.04µm. The microstructural characterization of the sintered samples was performed by a dual-beam Helios Nanolab 600i SEM equipped with Energy dispersive spectroscopy (EDS) and Electron backscattered diffraction (EBSD) detector. Phase identification was performed by X-ray diffraction (XRD) in a PANalytical diffractometer with a Cu K$\alpha_1$ radiation source. The XRD peaks obtained were analyzed using HighScore plus software. Vickers hardness tests were carried out with a load of 1Kgf for a period of 25s with a Shimizu hardness tester.

The tensile tests of all the samples were carried out at room temperature using a Kammrath and Weiss GmbH micro-tensile machine which can also be placed inside an SEM for in-situ testing. The tensile speed was $10^{-3}$ s$^{-1}$ and the samples were machined to a flat dog-bone shape with gauge dimensions of 4 x 1 x 1 mm. For each material, the average was calculated from two tests. Compression tests were performed on both the as-sintered (AS) and heat-treated samples at room temperature and at temperatures of 600°C, 700°C, and 750°C in an Instron 3384 machine equipped with a 30kN load cell. The sample dimensions were 2 x 2 x 4 mm. For each material, three tests were performed at a strain rate of $10^{-3}$/s and the average was calculated.

Table 1 : Proposed commodity alloys and their role in the target HEA

| Alloy | Size (d50) (µm) | Role | Ni | Fe | Cr | Mo | Co |
|---|---|---|---|---|---|---|---|
| | | | wt. (%) | | | | |
| Ni625* | 60 | Source of Ni, Cr, Fe and Mo | 61.54 | 5.35 | 25.26 | 5.6 | - |
| INVAR 36 | 3.8 | Source of Fe and Ni | 34.8 | 65.14 | - | - | - |
| CoCrF75 | 30 | Source of Co, Cr and Mo | 0.51 | 0.8 | 32.57 | 3.72 | 62.41 |
| 316L | 11.5 | Source of Fe, Cr, Ni and Mo | 11.82 | 67.77 | 18.9 | 1.51 | - |
| Fe49Ni | 15 | Source of Fe, Ni | 47.76 | 52.24 | - | - | - |

*Ni625 consists of ~3 wt% Nb

Table 2 : Proposed mixes of commodity powders used to develop different possible HEAs

| Alloy | wt. % | | | | | at. % | | | | |
|---|---|---|---|---|---|---|---|---|---|---|
| | Ni625 | INVAR 36 | CoCrF75 | 316L | Fe49Ni | Ni | Fe | Cr | Mo | Co |
| C1* | 20 | 38 | 42 | | | 25.21 | 26.92 | 20.7 | 1.61 | 25.56 |
| C2* | 28 | | 38 | 34 | | 21.01 | 25.03 | 28.56 | 2.1 | 23.28 |
| C3 | | | 48 | | 52 | 24.45 | 28.22 | 17.2 | 1.06 | 29.07 |

*C1 and C2 contain trace amounts of Nb

## 4. Results and Discussion

### 4.1 X-Ray Diffraction

The XRD peaks obtained for C1, C2 and C3 both in the as-sintered and heat treated state are shown in Fig. 2 (a), (b) and (c) respectively. C1 in the as-sintered state contains both FCC and HCP peaks, whereas after annealing at 1200°C for 24 hours, only FCC peaks remain. The same goes for C2 and C3, except that in C2 alloy sintered at 1100°C and C3 alloy sintered at 1000°C and 1100°C, no HCP peaks were detected in XRD but as shown in Fig S6, Fig S7 from the supplementary data and Fig. 6, HCP phases were present on the phase maps obtained by EBSD. The HCP phase is linked to the Co base powders and Co is highly concentrated in very specific areas in the as-sintered samples which makes it suitable to be detected by EBSD but not easily by XRD, thus, making the HCP peaks too weak. Both C2 and C3 alloy at both temperatures converted to a single FCC phase after annealing. The fact that all FCC peaks in all the alloys remain in the same position after annealing implies that there was no change in the lattice parameters.

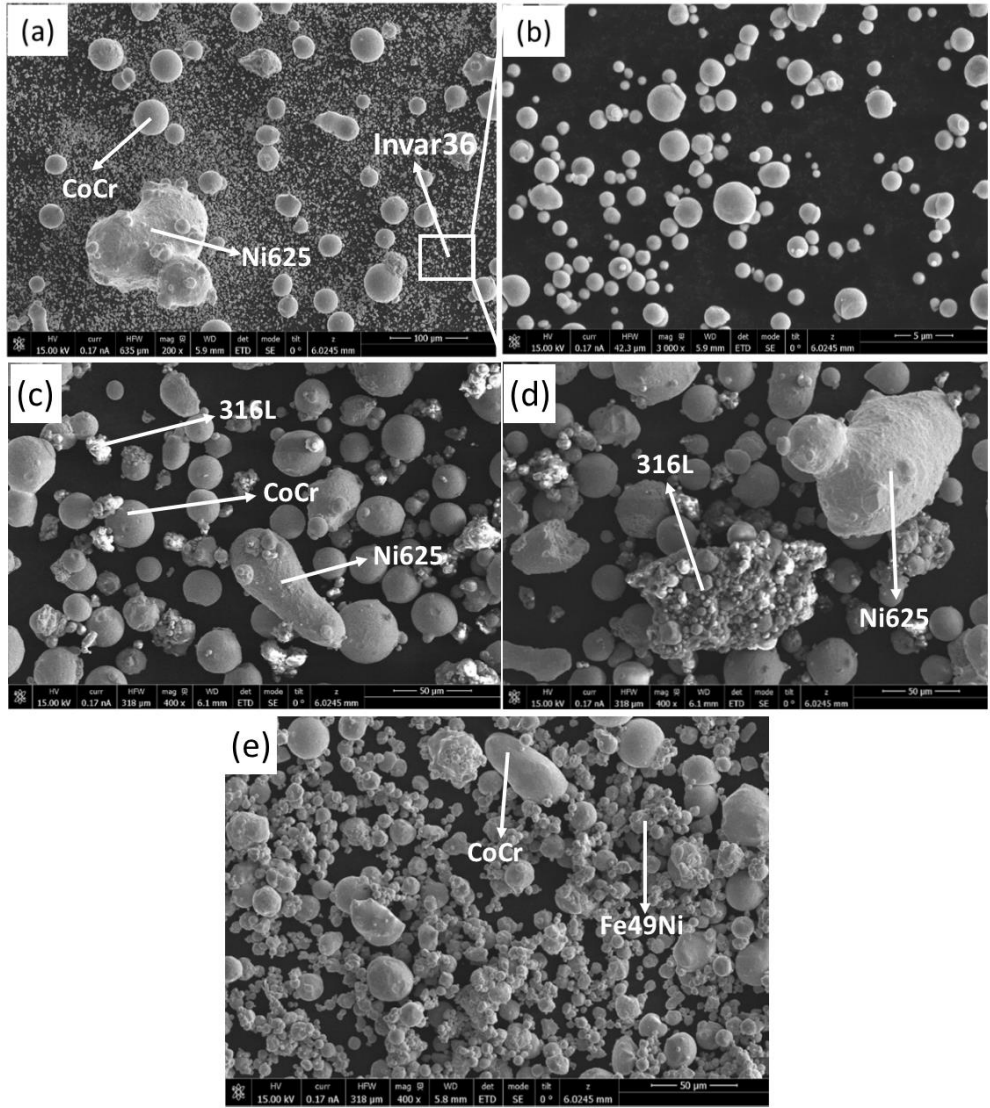

Fig. 1 : SEM images showing the powder morphology of (a) C1 mix; (b) magnified SEM image of Invar36 powders in the inset ; (c) and (d) powder morphology of C2 mix; (e) powder morphology of C3 mix

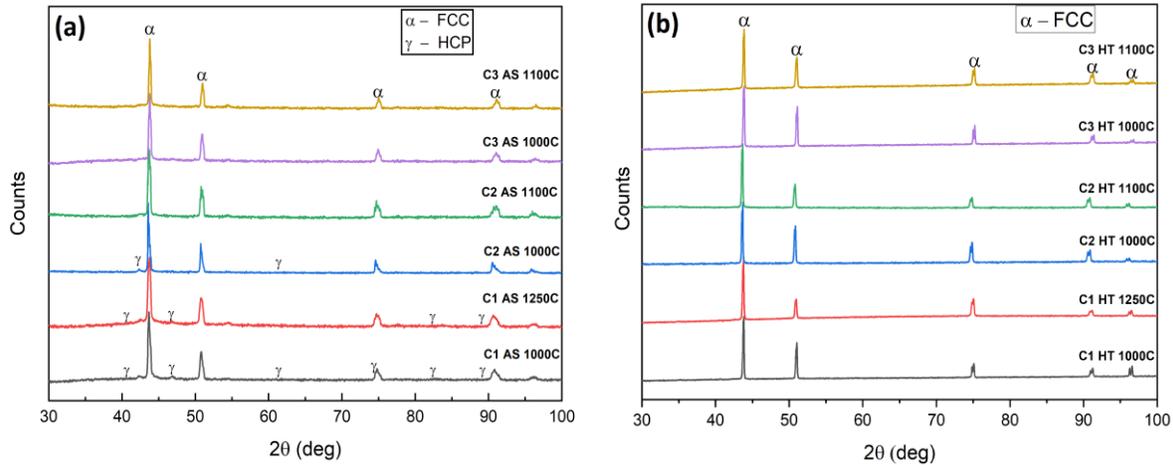

Fig. 2. XRD peaks of the alloys in the (a) as-sintered condition sintered at different temperatures and (b) after annealing at 1200°C for 24 hours

### 4.2 Microstructural analysis

The elemental distribution map obtained by EDS for C1 alloy in the as-sintered (sintered at 1000°C) and annealed state are shown in Fig. 3 and Fig. 4 respectively. The individual powder particles of C1, i.e., Ni625, Invar36, and CoCrF75 are clearly visible in the as-sintered state. The Invar36 powders formed the matrix here as its smaller particle size distribution enabled it to achieve higher sinterability. However, when the same alloy was annealed at 1200°C for 24 hours, the elemental distribution became homogenous, resulting in a typical HEA microstructure as shown in Fig. 4. The same pattern was followed for C2 and C3 alloys as well.

The microstructures of the as-sintered and annealed alloys sintered at 1000°C and 1100°C along with the porosity values are shown in Fig. S1 and Fig. S2 in the supplementary data. In the as-sintered microstructures, the individual powders can be seen whereas they are homogenized after annealing giving a typical HEA microstructure. In Fig. S1, it can be observed that the porosity decreases with higher sintering temperatures. The C2 mix of powderwasre not fully sintered well at the given parameters (1000°C, 50 MPa, 10 mins) due to the higher amount of large particle sized powders, which is Ni625, with a d50 of 60μm. However, the same mix of powders were properly sintered at a temperature of 1100°C.

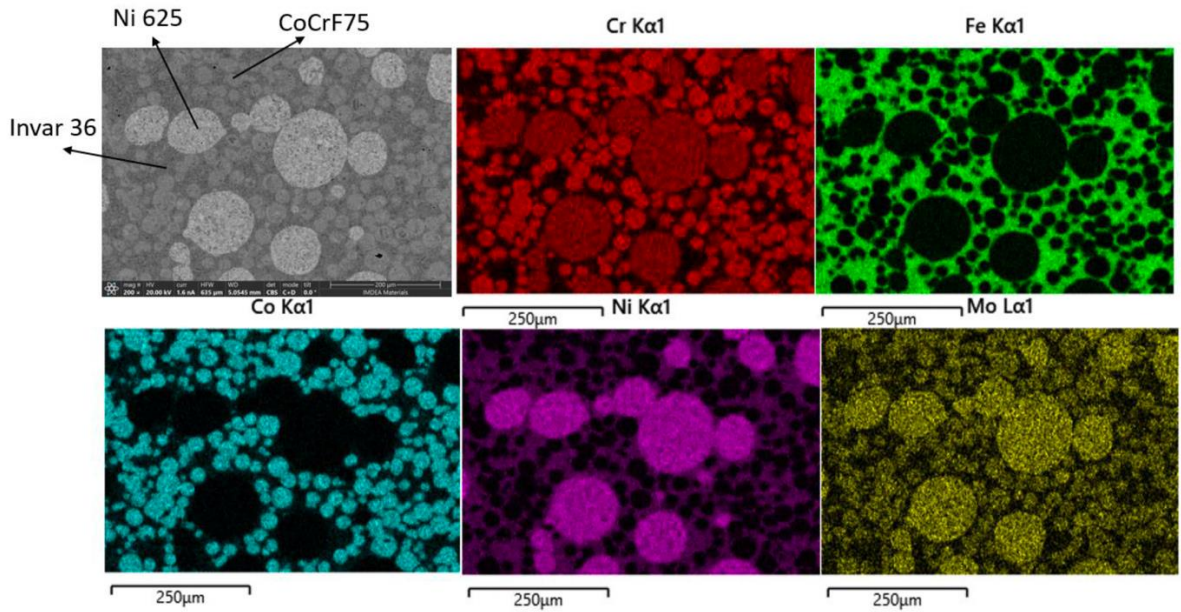

Fig. 3. Elemental distribution map of C1 alloy as-sintered via SPS at 1000°C

In highly localized regions, there were small white precipitates in C1 and C2 alloy after annealing but were absent in C3 alloy. A line scan of a precipitate is shown in Fig. S3 for C1 alloy sintered at 1000°C and annealed at 1200°C for 24 hours which shows that the precipitates are rich in Nb and Mo. As mentioned earlier, Nb is present in Ni625 alloy in small amounts which is mixed in C1 and C2. Due to its very low fraction in the microstructure, perhaps they were not identifiable by XRD. It is known that in Ni625 alloy, many intermetallic phases (like $Ni_3Nb$) and carbides (MC, $M_6C$ and $M_{23}C_6$) may precipitate after long-time annealing [16]–[19]. Also shown in Fig. S3 are some oxide inclusions which were confirmed by EDS showing chromium and oxygen peaks. As these were absent in the as-sintered alloys, it can be implied that they were picked up during the long-time annealing process. Regardless, these precipitates and oxides did not significantly influence the mechanical properties due to its low volume fraction as discussed in Section 4.3.2.

To study the phases after sintering and annealing, EBSD was performed on all the samples. The C1 alloy in the as-sintered state sintered at 1000°C consists of FCC, HCP and minor amount of BCC phases as shown in Fig. 5 (b). After annealing at 1200°C for 24 hours, it results in a single FCC phase as shown in Fig. 5 (d). The same C1 alloy when sintered at a higher temperature of 1250°C, consists of FCC but with much lesser amounts of HCP and BCC compared to when it was sintered at 1000°C as shown in Fig S4 (b). This is attributed to the fact that at higher temperatures, the atomic diffusion is faster which enables more homogeneity to occur. When this alloy was annealed at 1200°C for 24 hours, a single FCC phase was obtained as shown in Fig. S4 (d). The C2 alloy was sintered at 1000°C and 1100°C, referred to as C2_1000 and C2_1100 respectively. Both of them consisted of a majority of FCC phase and some amounts of BCC and HCP in the as-sintered condition as shown in Fig. S5 (b) and Fig. S6(b), respectively. However, after annealing, a single FCC phase was obtained in both the cases as shown in Fig. S5(d) for C2_1000 and Fig. S6(d) for C2_1100. The C3 alloy was also sintered at 1000°C and 1100°C. In both cases, the as-sintered samples consisted of FCC and HCP phases as shown in Fig. 6 (b) and Fig. S7 (b)**,** respectively**.** After annealing at 1200°C for 24 hours, the microstructure changed to a single FCC phase as shown in Fig. 6 (c) for C3_1000 and Fig. S7(d) for C3_1100. In all the alloys, after heat treatment, numerous annealing twins can be observed in the inverse pole figure maps suggesting low stacking fault energies as commonly found in recrystallized FCC metals [20].

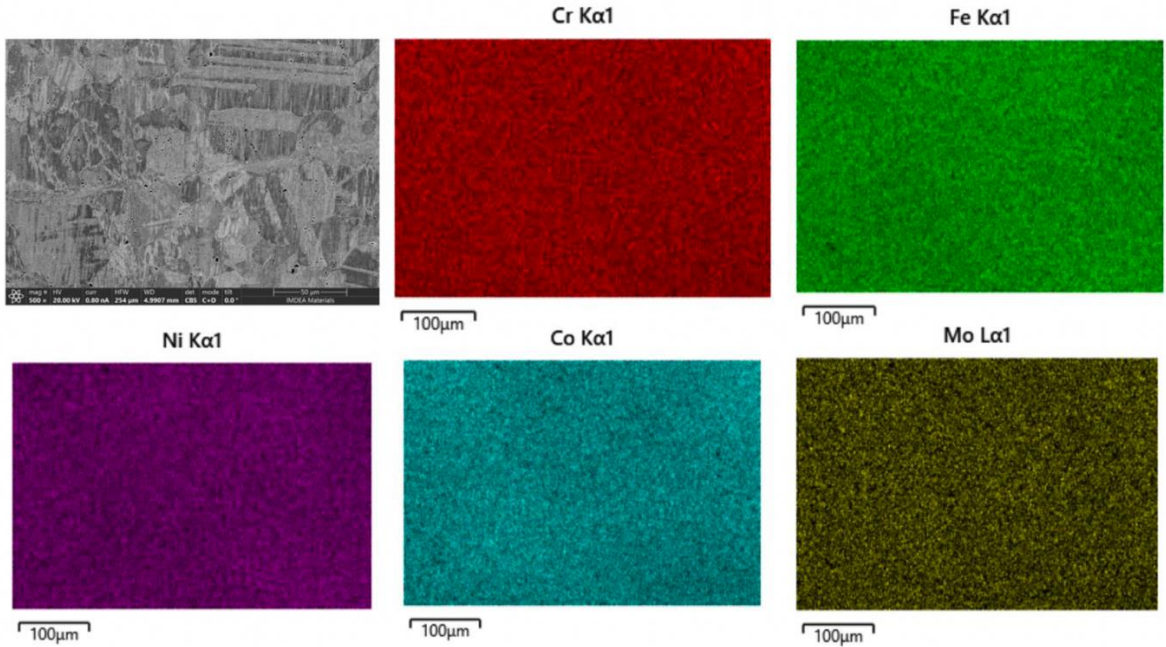

Fig. 4: Elemental distribution map of C1 alloy annealed for 24 hours

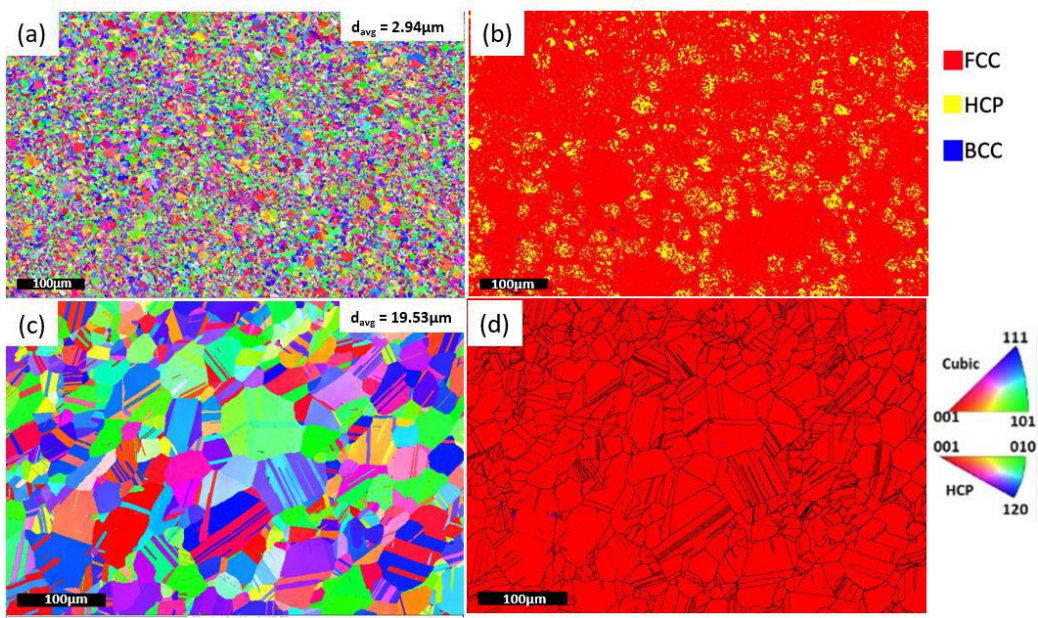

Fig. 5. IPF and phase map of C1 alloy sintered at 1000°C in the as-sintered state (a, b) and after annealing at 1200°C for 24hrs (c, d)

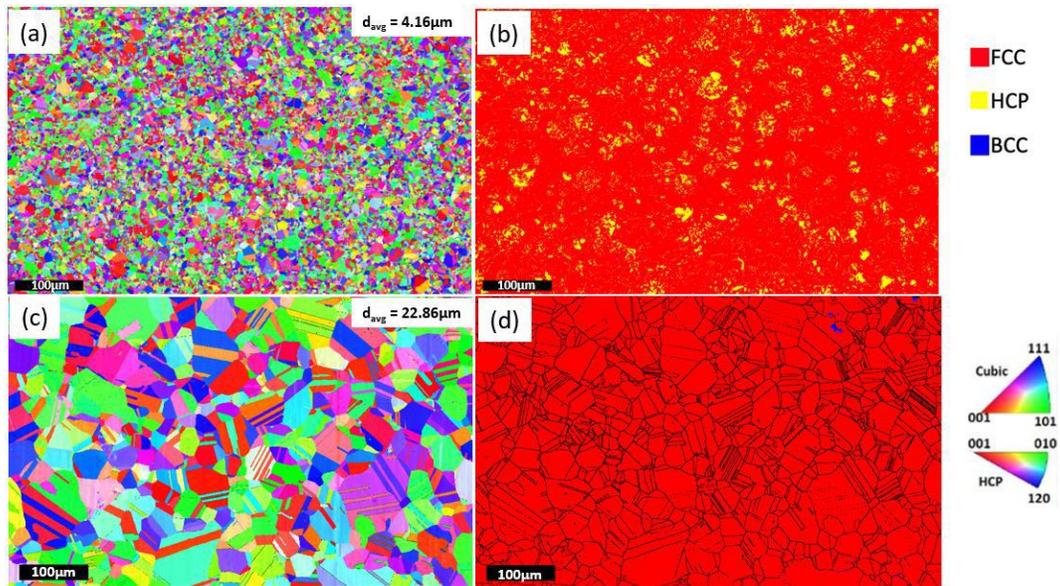

Fig. 6. IPF and phase map of C3 alloy sintered at 1000°C in the as-sintered state (a, b) and after annealing at 1200°C for 24hrs (c, d)

It is interesting to note that despite the high temperature and times used for annealing, the grain growth is only slight due to the very small grains obtained due to the fast heating rates and short dwell times in SPS. The grain sizes obtained after sintering and annealing in all the alloys are shown in Fig. 7. These grains are much smaller than the equivalent alloys obtained from ingot casting. For example, FeCoCrNiMo$_{0.1}$ had a grain size of 75µm when annealed at 1050°C for 60 minutes [21] which shows that powder metallurgy routes have great potential in improving the mechanical properties of new alloys.

## 4.3 Mechanical properties

### 4.3.1 Hardness

The vicker's hardness of the annealed alloys are shown in Fig. 8. The trend in hardness can be explained by the amount of Mo content. Hardness increases with Mo content as it aids in local lattice distortion effect in alloys [22]. Thus, C2 has the highest hardness followed by C1 and C3. These values are consistent with other similar PMHEAs like CoCrFeNi, where the obtained hardness was 194 Hv [23].

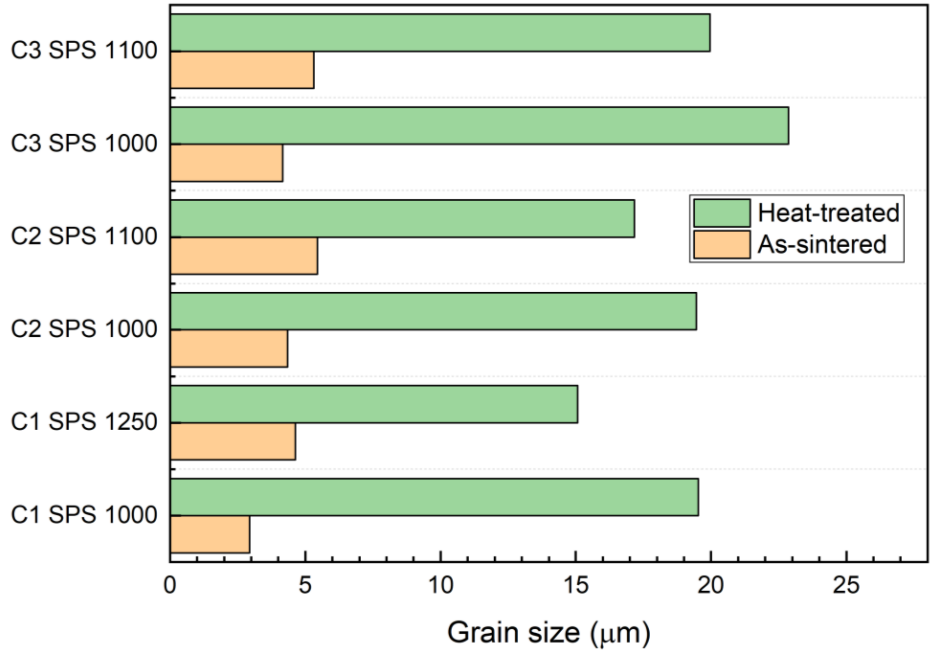

Fig. 7. Plot comparing the grain size of as-sintered and annealed samples obtained via SPS and FAHP

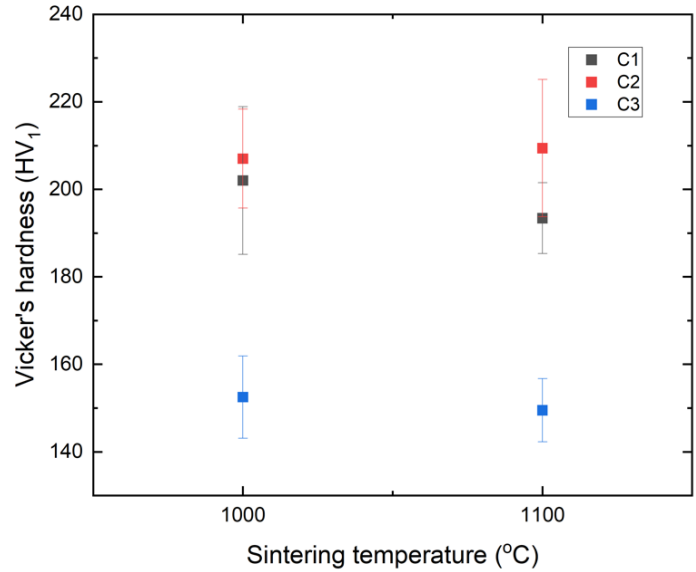

Fig. 8. Vicker's hardness values of C1, C2, and C3 alloys processed by SPS at various temperatures and annealed at 1200°C for 24 hours. It should be noted that the C1 alloy was sintered at 1250°C instead of 1100°C

### 4.3.2 Tensile properties

Micro-tensile tests were performed on the as-sintered and annealed samples. The as-sintered samples do not exactly represent HEAs as they are a mix of different powder particles and the composition is not uniform. But for comparative purposes, the tensile results were studied for the as-sintered samples also as shown in Fig. 9(a).

The tensile test results of annealed samples are shown in Fig. 9(b) and the summary of the mechanical properties of both as-sintered and annealed samples are shown in Table 3. The ultimate tensile strength (UTS) is lesser and the ductility is higher for all heat treated samples compared to the as-sintered ones due to the presence of a single FCC phase and larger grains in the former. All the alloys exhibit a reasonably high UTS and ductility for a single FCC phase, with C2 alloy exhibiting the maximum UTS of 712 MPa due to higher amounts of Mo. As mentioned earlier, the alloying of Mo causes local distortion in the FCC structure [22], which contributes to the solid solution strengthening.

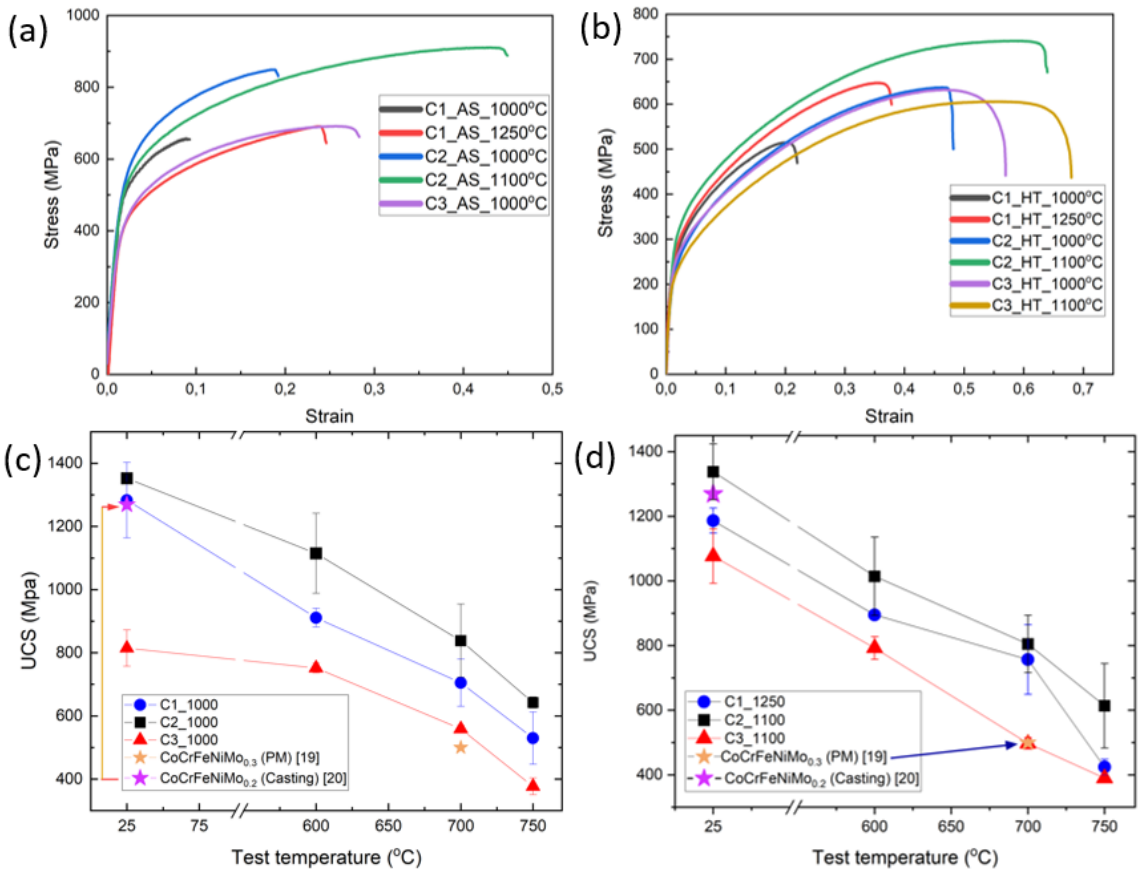

Fig. 9: Tensile stress-strain diagram of (a) As-sintered alloys (AS), (b) heat-treated alloys (HT), and ultimate compression strength values at various temperatures HT samples sintered at (c) 1000°C and (d) 1100°C. C1 alone was sintered at 1250°C instead of 1100°C.

Unfortunately, not a lot of work has been reported so far on tensile properties of PMHEAs, especially by SPS. Liu et al [22] investigated the tensile properties of Mo alloyed as-cast CoCrFeNi and reported a UTS of 479 MPa with 51.1% ductility for $CoCrFeNiMo_{0.1}$ and 589.6 MPa with 55.2% ductility for $CoCrFeNiMo_{0.2}$.

Similar FCC based alloys like CoCrFeNiMn had a yield strength of 200 MPa and a UTS of 600 MPa after arc-melting and hot rolling [24]. This shows that the same alloys developed in this work using powder metallurgy routes have better mechanical properties compared to their as-cast counterparts.

**Table 3: Tensile properties of C1, C2, and C3 alloys in the as-sintered (AS) and heat treated (HT) state**

| Alloys | Yield strength (MPa) | Ultimate tensile strength (MPa) | Ductility (%) |
|---|---|---|---|
| C1_AS_1000°C | 397.5 | 657.59 | 10.2 |
| C1_AS_1250°C | 311 | 691.07 | 25 |
| C2_AS_1000°C | 400 | 848.98 | 20 |
| C2_AS_1100°C | 418 | 894 | 42.75 |
| C3_AS_1000°C | 315 | 667.5 | 28 |
| C1_HT_1000°C | 322 | 663.75 | 28.75 |
| C1_HT_1250°C | 234 | 647.5 | 37.8 |
| C2_HT_1000°C | 190 | 637.4 | 48.2 |
| C2_HT_1100°C | 257.5 | 712 | 62 |
| C3_HT_1000°C | 200 | 595 | 57.5 |
| C3_HT_1100°C | 167.5 | 560 | 68 |
| CoCrFeNiMo$_{0.1}$[22] | 198.8 | 479 | 51.1 |
| CoCrFeNiMo$_{0.2}$[22] | 254.7 | 589.6 | 55.2 |
| CoCrFeNiMn[24] | 200 | 600 | 39 |

As seen in Table 3, in the as-sintered samples, the UTS increases with increasing sintering temperature as there is a shrinkage in porosity, even though there is a slight increase in the grain sizes. Among the heat treated samples, the C2 alloy shows an expected trend, where the YS and UTS increase with increasing sintering temperatures due to the reduction in porosity, but not in the case of C1 and C3. As mentioned earlier, C1 experienced melting when it was sintered at 1250°C, and produced a liquid phase which resulted in irregular porosities as shown in Fig. S8 and might have reduced the strength of the alloy. This is the same trend as seen from the values of vickers hardness as well where C1 sintered at 1250°C has lesser hardness than the one sintered at 1000°C. It should be noted that C3 alloy has the lowest YS and

UTS, due to its lower quantity of Mo and absence of Nb, which reduced the lattice distortion effect. This shows that Mo and Nb have a significant effect in strengthening HEAs even in small quantities.

Despite the lower ductility of C1_HT_1000°C, the fracture surfaces of all samples subjected to the tensile test shown in Fig. 10 present a typical micro-ductile fracture. All three samples exhibit micro-fracture with similar small dimples with large ductility signs. As shown in Fig. 10(d), some inclusions are present inside the larger dimples which upon EDS analysis showed two types of oxides, which were Mn-Cr rich and some Si-rich. The commodity powders used consists of trace amounts of Mn and Si which might have resulted in the formation of oxides after the long time annealing. However, no trends were observed between the oxide volume fraction and mechanical properties which shows that the oxides did not affect the mechanical properties significantly. Moreover, no Nb and Mo rich precipitates were found on the fracture surfaces of C1 and C2 alloys again indicating their insiginificant impact on the mechanical properties.

### 4.3.3 Hot compression test

The ultimate compression strength of the annealed alloys tested at various temperatures up to 750°C obtained by hot compression test is shown in Fig. 9(c) for samples sintered at 1000°C and Fig. 9 (d) for samples sintered at 1100°C. The summary of the yield strength (YS), ultimate compression strength (UCS) and ductility of both as-sintered (AS) and annealed samples (HT) are presented in Table S1 in the supplementary data. The AS alloys in general exhibit better strength and lesser ductility than HT alloys due to the presence of harder and stronger HCP and BCC phases, and the fine grain sizes obtained in the former.

As expected, in all the alloys, the UCS decreases with increasing temperatures as shown in Fig. 9 (c) and (d) due to the dynamic softening that occurs at higher temperatures due to the improved thermal activation process [25]. Similar to the tensile test results, the C2 alloy performs better at all temperatures compared to C1 and C3 due to the higher amount of Mo and Cr, which aids in solid solution strengthening. It is to be noted that all the heat treated samples experienced no fracture at any of the test temperatures even after the strain reached more than 40% indicating a remarkable ductility of the FCC phase. For example, at RT, the C2_HT alloy sintered at 1000°C exhibited a UCS of 1352.5 MPa with a ductility >40%. The same alloy at 750°C, had a UCS of 642.3 MPa with a ductility of >44%. Comparing similar alloys reported in the literature, $CoCrFeNiMo_{0.3}$ developed by casting, showed a UCS of 1269 MPa with a ductility of 58% at room temperature [26]. The same alloy when developed by PM route had a UCS of 500 MPa at 700°C [27] which is lower than that of the HEAs studied in this work.

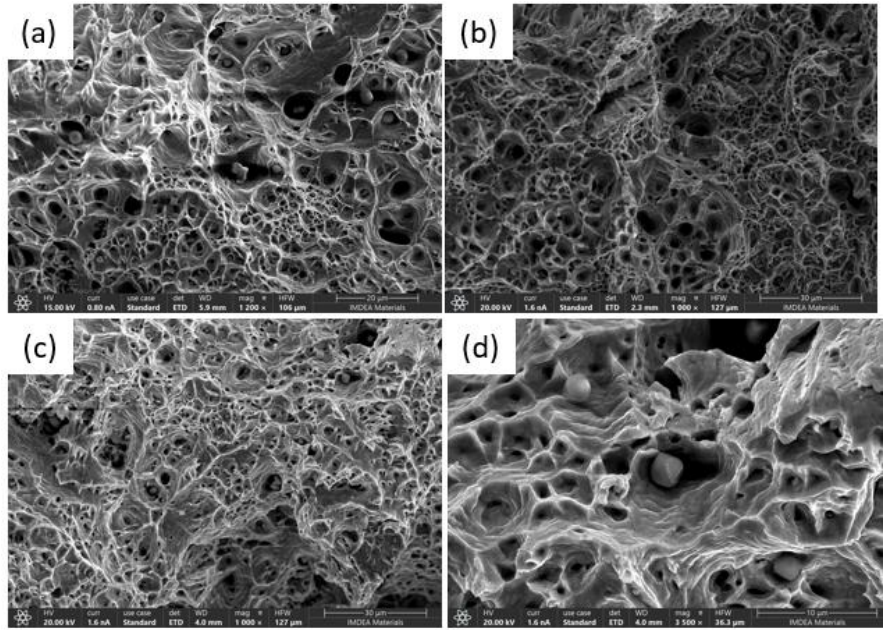

Fig. 10: SEM images of fracture surfaces of (a) C1, (b) C2, (c) C3, (d) C3 at higher magnification, sintered at 1000°C and annealed at 1200°C for 24 hours subjected to tensile testing

## 5. Cost analysis

A cost analysis was conducted to compare the fabrication route by pure elemental powders and commercial commodity powders. Based on the quote by raw materials provider Hunan Fushel Technology Limited [28], a bar chart is shown in Fig. 11, comparing the cost of C2 alloy. It can be seen that by using commercial powders, there is a 20% reduction in costs than when using pure elemental powders. In addition, it should be noted that when using elemental powders, there are added costs of pre-alloying steps like mechanical alloying or gas atomization, which will add to the total manufacturing cost and time. Hence, this method of using commodity powders offers a feasible way for the commercialization and industrial implementation of HEAs.

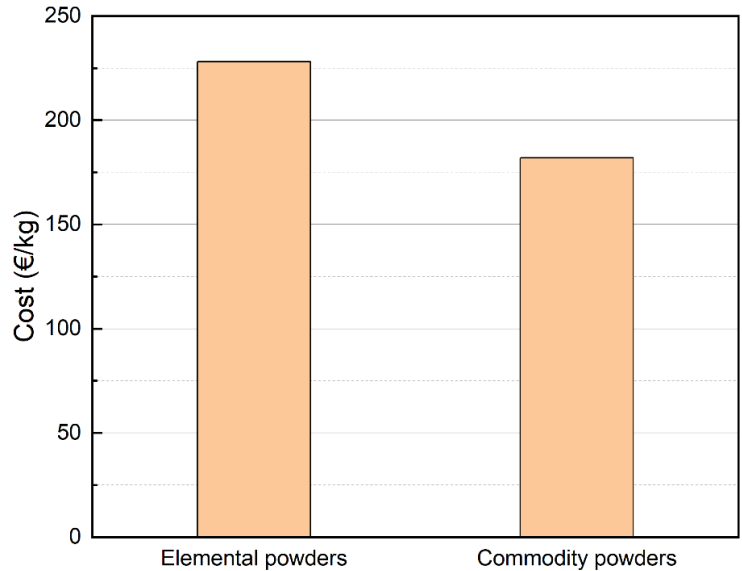

Fig. 11: Cost comparison of manufacturing C2 alloy using elemental powders vs commodity powders

# 6. Conclusions

The objective of this work was to explore the possibility of developing HEAs using commercial commodity powders which are readily available in the market in bulk quantities. Bulk samples of HEAs using a simple blend of commodity powders namely, Ni 625, Invar36, CoCrF75, 316L, and Fe49Ni were fabricated by SPS at 1000°C and 1100°C. Post-sintering, the microstructure consisted of individual alloy powder particles with multiple phases including FCC, HCP and BCC and after an annealing treatment at 1200°C for 24 hours, it gave rise to a typical HEA microstructure with an FCC phase.

1. Three different non-equiatomic HEAs, $Co_{25.56}Cr_{20.7}Fe_{26.92}Ni_{25.2}Mo_{1.6}$, $Co_{23.28}Cr_{28.57}Fe_{25.03}Ni_{21.01}Mo_{2.1}$, $Co_{29.07}Cr_{17.2}Fe_{28.22}Ni_{24.45}Mo_{1.06}$, labelled C1, C2 and C3 respectively were prepared by blending different combinations of the commercial powders. Both the as-sintered and annealed samples underwent hot compression tests from room temperature to 750°C and micro-tensile tests at room temperature. The as-sintered samples had better strength compared to the annealed samples due to the presence of HCP and BCC phases. The best mechanical properties, both tensile and compressive, from the annealed samples which are the actual HEAs, were exhibited by C2 alloy sintered at 1100°C due to the higher amount of Mo. It had a tensile YS of 257.5 MPa and a UTS of 712 MPa with a ductility of 62%. Also, it had a compressive YS of 330 MPa and a UCS of 1352.5 MPa at room temperature. Moreover, the same alloy had a YS of 142 MPa and a UCS of 642.3 MPa at 750°C. The annealed samples did not undergo any fracture during compression test even after reaching a strain of 45%, indicating remarkable ductility of the FCC HEAs. These values are higher than that found in the literature for similar HEAs obtained by as-cast methods and are also compatible with those obtained by powder metallurgy routes using fully pre-alloyed powders.

2. This work proves that HEAs can be developed by these commodity powders without the need for pre-alloying processes like mechanical alloying, thus avoiding contamination and paving way for an efficient, cost-effective and flexible way for industrial implementation and commercialization. For instance, this method of fabrication saved around 20% of the manufacturing cost for one of the alloys. So, the use of commodity alloy scraps as raw materials to manufacture HEAs using both casting and PM routes is a viable alternative, avoiding the direct use of critical metals like Co and Ni, due to both its toxicity and availability.

## Acknowledgements

The authors would like to thank VDM Metals for providing the powders for free. This work was funded by IMDEA Materials Institute, Madrid, Spain.

## Supplementary data

A supplementary data is attached for this article.

## Data Availability

Data will be made available on request.

**Declaration of interests**

☒ The authors declare that they have no known competing financial interests or personal relationships that could have appeared to influence the work reported in this paper.

☐The authors declare the following financial interests/personal relationships which may be considered as potential competing interests:

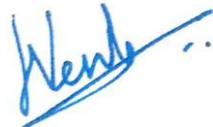

04/02/2023

# A novel and sustainable method to develop non-equiatomic CoCrFeNiMo$_x$ high entropy alloys via spark plasma sintering using commercial commodity powders and evaluation of its mechanical behaviour

S Venkatesh Kumaran[1,2], Dariusz Garbiec[3], José Manuel Torralba[1,2]

1. IMDEA Materials Institute, Madrid, Spain, 28906
2. Universidad Carlos III de Madrid, Leganes, Spain, 28911
3. Metal Forming Institute, 14 Jana Pawla II Street, 61-139 Poznan, Poland

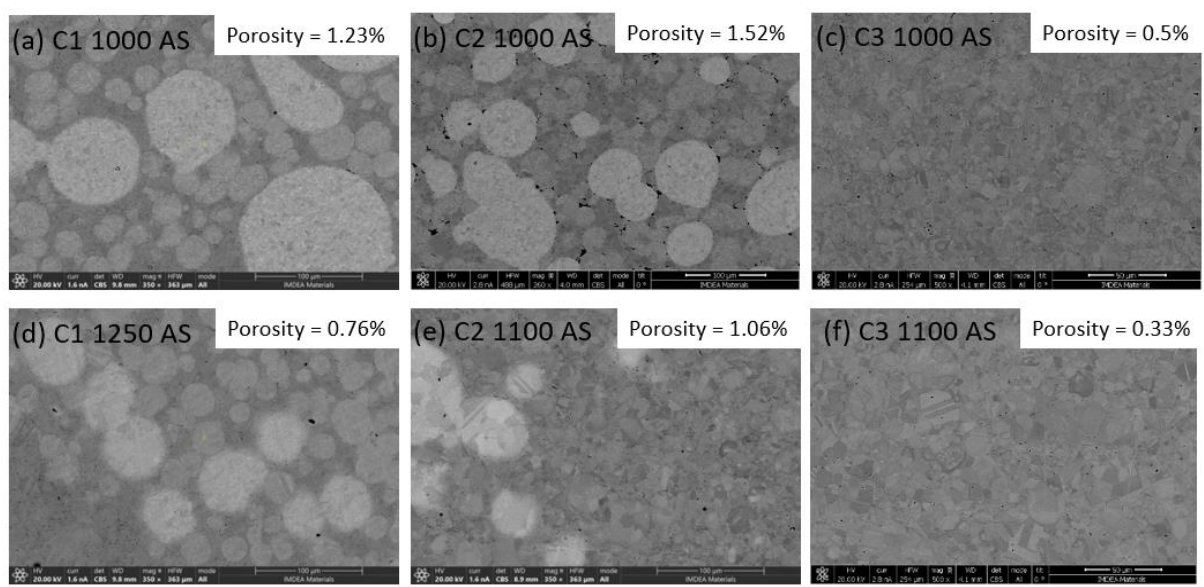

Fig. S 1: SEM images of as-sintered samples via SPS; C1 sintered at (a) 1000°C, (d) 1250°C, and C2 sintered at (b) 1000°C, (e) 1100°C, and C3 sintered at (c) 1000°C, (f) 1100°C

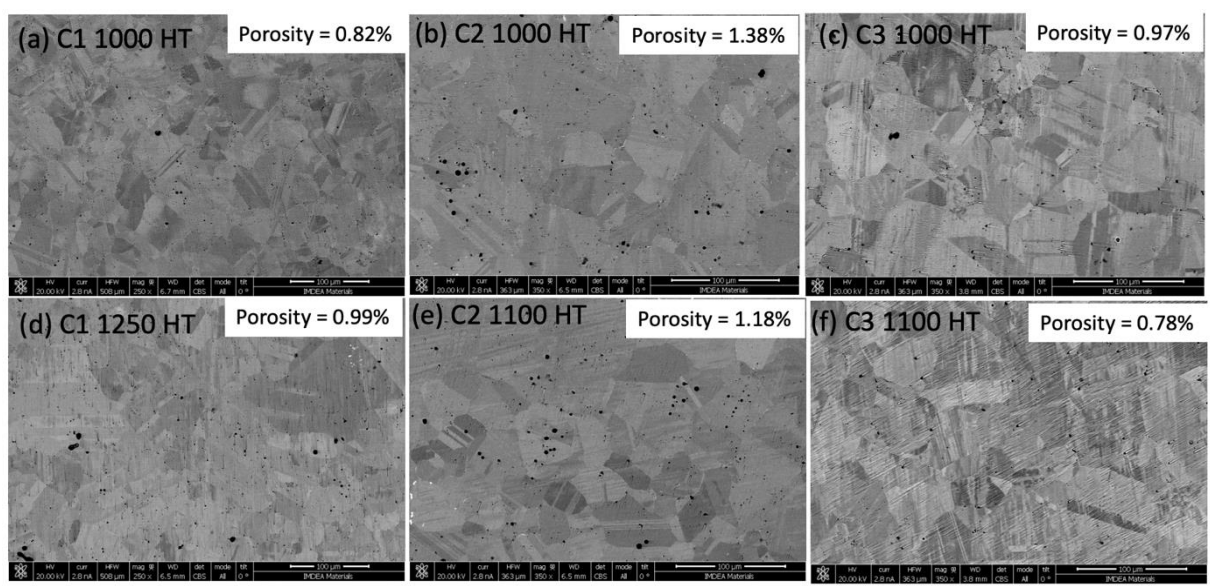

Fig. S 2: SEM images of samples sintered at various temperatures via SPS and annealed at 1200°C for 24 hours; C1 sintered at (a) 1000°C, (d) 1250°C, and C2 sintered at (b) 1000°C, (e) 1100°C, and C3 sintered at (c) 1000°C, (f) 1100°C

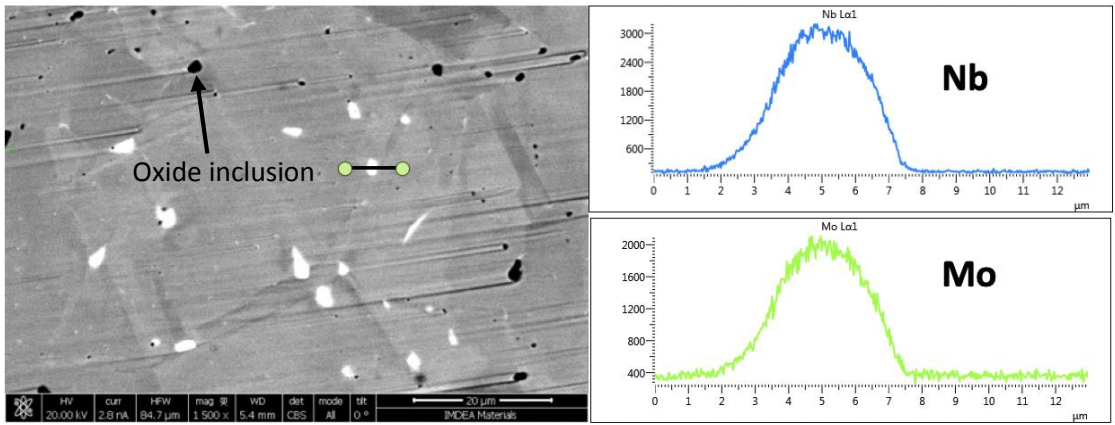

Fig. S 3: SEM BSE image of C1 alloy sintered at 1000°C and annealed at 1200°C for 24 hours with a line scan on a precipitate showing the distribution of Nb and Mo. The black arrow points at an oxide inclusion

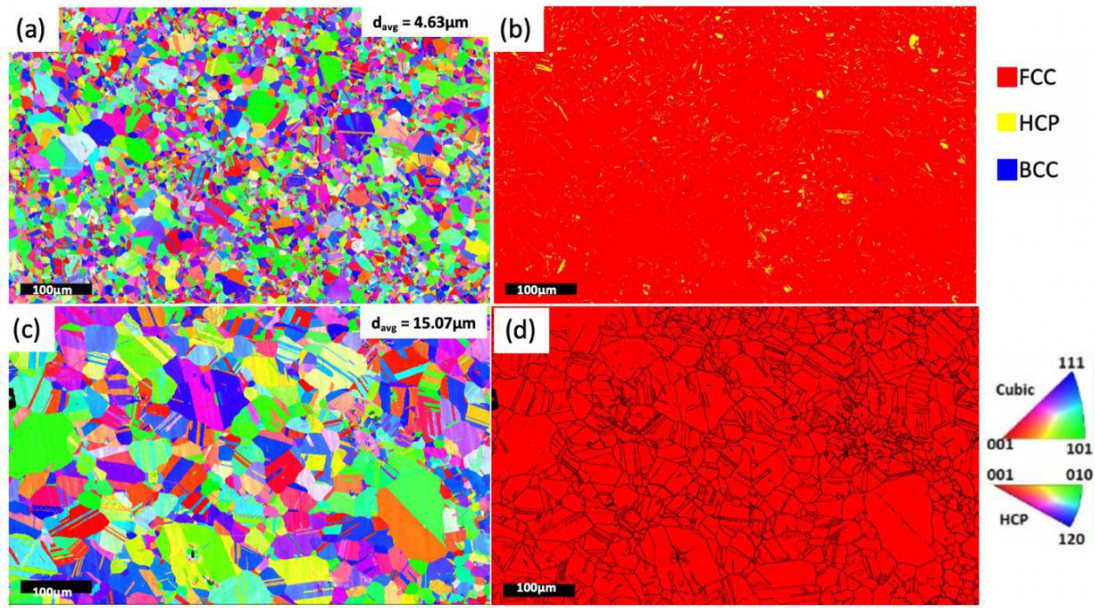

Fig. S 4: IPF and phase map of C1 alloy sintered at 1250°C in the as-sintered state (a, b) and after annealing at 1200°C for 24hrs (c, d)

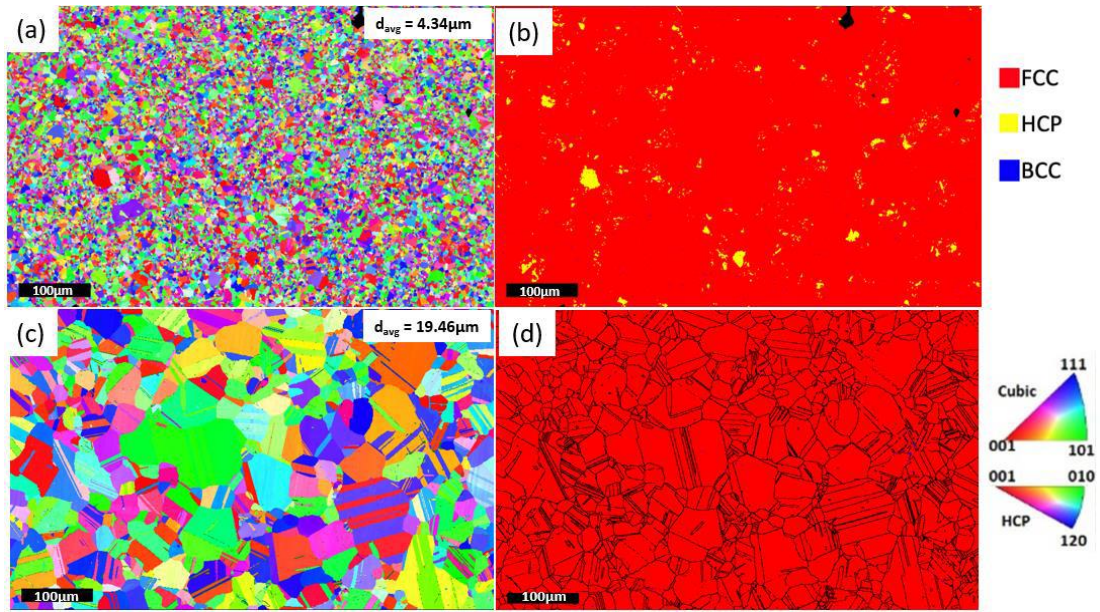

Fig. S 5. IPF and phase map of C2 alloy sintered at 1000°C in the as-sintered state (a, b) and after annealing at 1200°C for 24hrs (c, d)

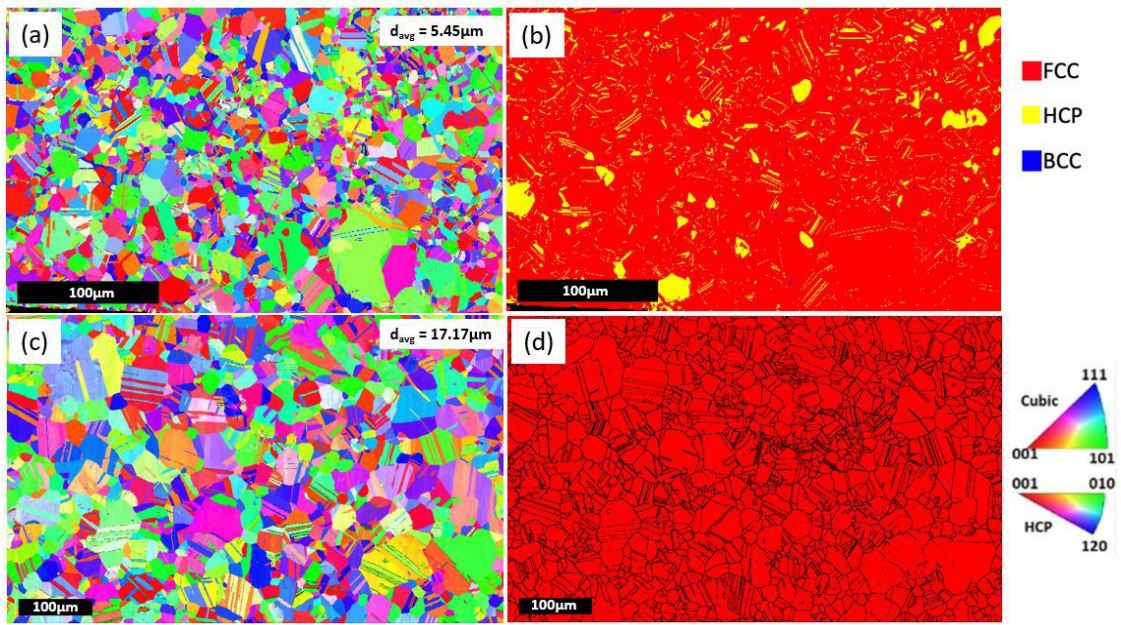

Fig. S 6. IPF and phase map of C2 alloy sintered at 1100°C in the as-sintered state (a, b) and after annealing at 1200°C for 24hrs (c, d)

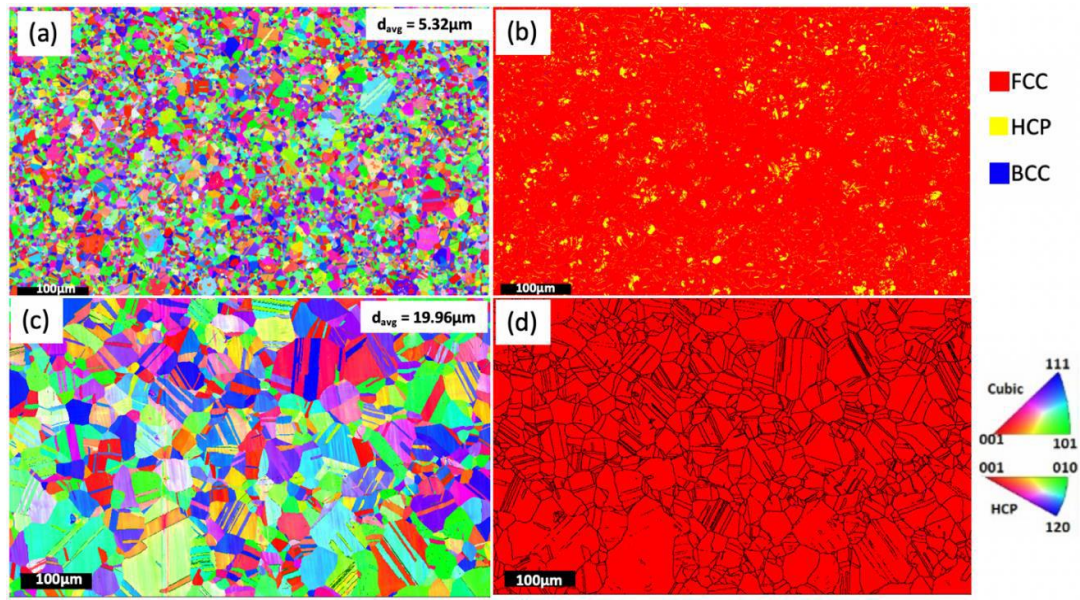

Fig. S 7. IPF and phase map of C3 alloy sintered at 1100°C in the as-sintered state (a, b) and after annealing at 1200°C for 24 hrs (c, d)

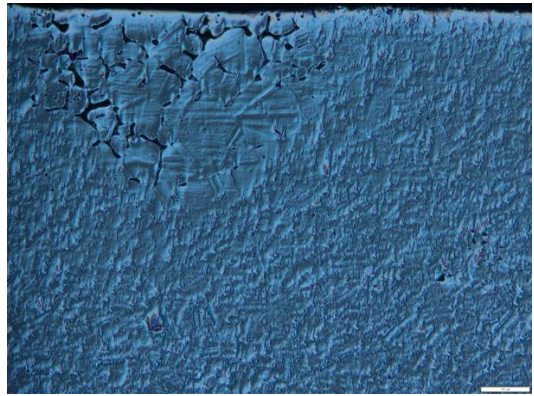

Fig. S 8. Optical microscope images of C1 alloy sintered at 1250°C and annealed at 1200°C for 24 hours

**Table S1: Results of hot compression properties of C1, C2 and C3 alloys in the as-sintered (AS) and heat-treated (HT) condition from room temperature to 750°C**

| Temperature | Property | C1 1000°C AS | C1 1250°C AS | C2 1000°C AS | C2 1100°C AS | C3 1000°C AS | C3 1100°C AS | C1 1000°C HT | C1 1250°C HT | C2 1000°C HT | C3 1000°C HT | C3 1000°C HT |
|---|---|---|---|---|---|---|---|---|---|---|---|---|
| RT | YS (MPa) | 370 | 268.8 | 572.3 | 506 | 501.7 | 335 | 310 | 211 | 330 | 205 | 183.6 |
|  | UTS (MPa) | 1242.5 | 1351.3 | 1547 | 1847.7 | 1134.4 | 1184 | 1224 | 1202.5 | 1352.5 | 871.5 | 1076.6 |
|  | e (%) | 31.5 | 33.7 | 31 | 37 | 35.8 | 39.2 | >44 | >42 | >41 | >39 | >39 |
| 600 | YS (MPa) | 310 | 335.8 | 394 | 320 | 277.6 | 210 | 120 | 190 | 129 | 163.6 | 134 |
|  | UTS (MPa) | 715.7 | 780.5 | 1039.2 | 1042.3 | 634 | 697 | 911 | 895 | 1115 | 751.6 | 792.3 |
|  | e (%) | 20.1 | 28.45 | 25.6 | 30 | 29.9 | 35.2 | >41.6 | >37.6 | >39.3 | >35.5 | >34.7 |
| 700 | YS (MPa) | 272.5 | 211.5 | 353 | 211.09 | 227.6 | 188.9 | 122.5 | 127.3 | 132 | 113.5 | 145.8 |
|  | UTS (MPa) | 525.2 | 663.4 | 802 | 806.3 | 446 | 503 | 705 | 683 | 757.5 | 549.7 | 497.4 |
|  | e (%) | 28.1 | 35.85 | 35.9 | 42.75 | 37 | 38.5 | >41.7 | >39.3 | >37.9 | >35.5 | >36.8 |
| 750 | YS (MPa) | 213 | 188 | 300 | 274.6 | 166 | 160.5 | 164 | 132.2 | 142 | 90 | 120.5 |
|  | UTS (MPa) | 352.5 | 410.8 | 488.7 | >596 | 265 | 313.8 | 530 | 424 | 642.2 | 376.6 | 390 |
|  | e (%) | 31 | 31.4 | 45 | >37.7 | 34.8 | 43.7 | >37.5 | >34 | >44.1 | >35.9 | >33.7 |